\documentclass[twocol]{ametsocV6.1}

\usepackage{graphicx,amsmath}

\title{Distilling Machine Learning's Added Value: Pareto Fronts in Atmospheric Applications}

\authors{Tom Beucler\aff{a,b}\correspondingauthor{Tom Beucler, tom.beucler@unil.ch}, Arthur Grundner\aff{c}, Sara Shamekh\aff{d}, Peter Ukkonen\aff{e}, Matthew Chantry\aff{f}, Ryan Lagerquist\aff{g,h}}

\affiliation{\aff{a}{Faculty of Geosciences and Environment, University of Lausanne, Lausanne, VD, Switzerland}\\  
\aff{b}{Expertise Center for Climate Extremes, University of Lausanne, Lausanne, VD, Switzerland}\\  
\aff{c}{Deutsches Zentrum für Luft- und Raumfahrt, Institut für Physik der Atmosphäre, Oberpfaffenhofen, Germany}\\  
\aff{d}{Courant Institute of Mathematical Sciences, New York University, New York,
NY, USA}\\   
\aff{e}{Department of Physics, University of Oxford, Oxford, United Kingdom}\\   
\aff{f}{European Centre for Medium-Range Weather Forecasts, Reading, United Kingdom}\\   
\aff{g}{Cooperative Institute for Research in the Atmosphere (CIRA), Colorado State University, Fort Collins, CO, USA}\\   
\aff{h}{National Oceanic and Atmospheric Administration (NOAA) Global Systems Laboratory (GSL), Boulder, CO, USA}}

\abstract{The added value of machine learning for weather and climate applications is measurable through performance metrics, but explaining it remains challenging, particularly for large deep learning models. Inspired by climate model hierarchies, we propose that a full hierarchy of Pareto-optimal models, defined within an appropriately determined error-complexity plane, can guide model development and help understand the models' added value. We demonstrate the use of Pareto fronts in atmospheric physics through three sample applications, with hierarchies ranging from semi-empirical models with minimal parameters (simplest) to deep learning algorithms (most complex). First, in cloud cover parameterization, we find that neural networks identify nonlinear relationships between cloud cover and its thermodynamic environment, and assimilate previously neglected features such as vertical gradients in relative humidity that improve the representation of low cloud cover. This added value is condensed into a ten-parameter equation that rivals deep learning models. Second, we establish a machine learning model hierarchy for emulating shortwave radiative transfer, distilling the importance of bidirectional vertical connectivity for accurately representing absorption and scattering, especially for multiple cloud layers. Third, we emphasize the importance of convective organization information when modeling the relationship between tropical precipitation and its surrounding environment. We discuss the added value of temporal memory when high-resolution spatial information is unavailable, with implications for precipitation parameterization. Therefore, by comparing data-driven models directly with existing schemes using Pareto optimality, we promote process understanding by hierarchically unveiling system complexity, with the hope of improving the trustworthiness of machine learning models in atmospheric applications.}

\begin{document}

\maketitle

%
%
%
\statement
As machine learning permeates the geosciences, it becomes urgent to identify improvements in process representation and prediction that challenge our understanding of the underlying system. We show that Pareto-optimal hierarchies transparently distill the added value of new algorithms in three atmospheric physics applications, providing a timely complement to post-hoc explainable artificial intelligence tools.

%
%
%

%
\section{Introduction}


Will recent advancements in machine learning (ML) lead to enduring new knowledge in atmospheric science? While the added value of ML for weather and climate applications can be measured using performance metrics, it often remains challenging to understand. Taking advancements in data-driven, medium-range weather forecasting \citep{ben2023rise} as an example, increasing reliance on complex architectures makes state-of-the-art models difficult to interpret. 

\cite{weyn2019can,weyn2020improving} used convolutional neural networks (CNN) with approximately 200k and 700k learned parameters to produce global forecasts that outperformed climatology, persistence, and a low-resolution numerical weather prediction (NWP) model for lead times smaller than one week. Following the success of early approaches, \cite{rasp2020weatherbench} developed a benchmark dataset for data-driven weather forecasting, which facilitated the objective assessment of rapid developments \citep[e.g.,][]{clare2021combining,scher2021ensemble}. Since then, for data-driven, medium-range forecasting, \cite{rasp2021data} trained a $\approx 6.3$M-parameter deep residual CNN, \cite{keisler2022forecasting} trained a $\approx 6.7$M-parameter graph neural network (GNN), and \cite{pathak2022fourcastnet} trained a $\approx 75$M-parameter emulator combining transformers with Fourier neural operators. Recently, deep learning models started rivaling state-of-the-art, high-resolution, deterministic NWP models: \cite{lam2022graphcast} via combined GNNs totaling $\approx37$M parameters, \cite{bi2022pangu} via a $\approx256$M-parameter Earth-like transformer, and \cite{lang2024aifs} via a $\approx256$M-parameter graph and transformer model. It is hard to pinpoint what makes these models so successful, even with modern explainable artificial intelligence \citep[XAI;][]{buhrmester2021analysis,das2020opportunities} tools, given that XAI requires certain assumptions to be satisfied \citep{mamalakis2022investigating,mamalakis2023carefully} and involves choosing which samples to investigate, which is challenging for large models and datasets. 

The growing complexity of data-driven models for weather applications shares similarities with the development of general circulation models (GCM) that followed the first comprehensive assessment of global climate change due to carbon dioxide \citep{charney1979carbon}. Unlike data-driven weather prediction, where reducing forecast errors could warrant increased complexity, GCMs have traditionally been created to not simply project but also comprehend climate changes \citep{held2005gap,balaji2022general}. This implies that any additional complexity in an Earth system model should be well-justified, motivating \textit{climate model hierarchies} that aim to connect our fundamental understanding with model prediction \citep[e.g.,][]{mansfield2023updates,robertson2000solving, bony2013carbon,jeevanjee2017perspective,maher2019model,balaji2021climbing}. 

Inspired by climate model hierarchies, we here show that modern optimization tools help systematically generate \textit{data-driven model hierarchies} to model and understand climate processes for which we have reliable data. These hierarchies can: (1) guide the development of data-driven models that optimally balance simplicity and accuracy; and (2) unveil the role of each complexity unit, furthering process understanding by distilling the added value of ML for the atmospheric application of interest. 

In this study, we showcase the advantages of considering a hierarchy of models with varying error and complexity, as opposed to focusing on a single fitted model \citep{fisher2019all}. After formulating Pareto-optimal model hierarchies (Sec.~\ref{sec:Pareto_Optimal_hierarchy}) and categorizing the added value of ML into four categories (Sec.~\ref{sec:Added_Value_ML}), we apply our approach to three atmospheric processes relevant for weather and climate predictions (Sec.~\ref{sec:Weather_Climate}) to distill the added value of recently developed deep learning frameworks before concluding (Sec.~\ref{sec:Conclusion}). 

\section{Pareto-Optimal Model Hierarchies\label{sec:Pareto_Optimal_hierarchy}}

 
\subsection{Pareto Optimality}

In multi-objective optimization, Pareto optimality represents a solutions set that cannot be improved upon in one criterion without worsening another criterion \citep[e.g.,][]{censor1977pareto,miettinen1999nonlinear}. The first step is to define a set of $n$ real-valued model evaluation metrics $\mathcal{E} = \left\{ {\cal E}_{i}\right\} _{i=1}^n$ that we wish to minimize (e.g., error, complexity). We call a model $M_{\mathrm{opt}} $ \textit{Pareto-optimal} w.r.t. these metrics and w.r.t. a model family if there is no model in that family that strictly outperforms $M_{\mathrm{opt}}$ in one metric while maintaining at least the same performance in all other metrics \citep[e.g.,][]{lin2019pareto}. The \textit{Pareto front} (PF) is the set of all Pareto-optimal models, which can be defined using logical statements:

\begin{equation}
\mathrm{PF}_{{\cal E}}=\left\{ M_{\mathrm{opt}}\ |\ \nexists M\ \mathrm{s.t.\ }
\begin{cases}
\forall i\:{\cal E}_{i}\left(M\right)\leq {\cal E}_{i}\left(M_{\mathrm{opt}}\right)\\
\exists j\:{\cal E}_{j}\left(M\right)<{\cal E}_{j}\left(M_{\mathrm{opt}}\right)
\end{cases} \right\}.
\end{equation}

Intuitively, when we select a model from the Pareto front, any attempt to switch to a different model would mean sacrificing the quality of at least one evaluation metric. Conversely, a model that can be replaced without compromising any evaluation metrics is described as \textit{Pareto-dominated}. Henceforth, we derive Pareto fronts empirically from the available data and the subset of models $\cal M$ considered. These empirical Pareto fronts are denoted $\mathrm{PF}_{{\cal E},{\cal M}}$, acknowledging that the ``true Pareto front'' $\mathrm{PF}_{{\cal E}}$ is generally intractable. In practice, we often seek to balance evaluation metrics measuring error and complexity, which we discuss in the next subsections. 



\subsection{Error}
We emphasize the importance of \textit{holistic} evaluation, which employs several error metrics with different behaviors: Traditional regression or classification metrics, distributional distances, spectral metrics, probabilistic scoring rules, reliability diagrams \citep{haynes2023creating}, causal evaluation \citep{nowack2020causal}, etc. To facilitate the use of Pareto-optimal hierarchies, we recommend prioritizing proper scores, whose expectation is optimal if and only if the model represents the true target distribution \citep{brocker2009reliability}. For simplicity's sake, we will employ mean squared error (MSE) and its square root (RMSE) as our primary error metric for our study's applications. We make this choice because MSE is a proper score for deterministic predictions that can be efficiently optimized, while recognizing MSE's inherent limitations for non-normally distributed targets. 

\subsection{Complexity}
To our knowledge, there are no universally accepted metrics for quantifying the complexity of data-driven models within the geosciences. In statistical learning, various complexity metrics, such as Rademacher complexity \citep[e.g.,][]{bartlett2005local}, rely on dataset characteristics, whereas others, like the VC dimension \citep{vapnik2015uniform}, solely depend on algorithmic attributes. Here, we predominantly focus on two metrics that can be readily calculated from model attributes: the number of trainable parameters ($N_\mathrm{param}$) and the number of features ($N_\mathrm{features}$). We choose these metrics due to their simplicity and versatility across the broad spectrum of models considered in our study \citep{balaji2017cpmip}. As we will confirm empirically in Sec.~\ref{sec:Weather_Climate}, $N_\mathrm{param}$ and $N_\mathrm{features}$ can be used as (very) approximate proxies for generalizability, as models with fewer trainable parameters and features with more stable distributions tend to result in superior generalizability. We defer the exploration of additional complexity metrics, such as the number of floating point operations (FLOP) or Multiply-Accumulate Operations (MACs), to future work.

Equipped with these definitions, we can now ask: Why, along the Pareto front in a well defined error-complexity plane, does increasing complexity result in better performance? In the following section, we hence leverage Pareto optimality to distill machine learning's added value.

\section{The Distillable Value of Machine Learning \label{sec:Added_Value_ML}}

Amid rapid progress in optimization, machine learning architectures, and data handling, it can be challenging to distinguish long-lasting progress in modeling from small improvements in model error. We postulate that progress is more likely to be replicable if we can \textit{explain} how added model complexity improves performance in simple terms. Based on this postulate, we propose simple definitions to categorize a model's added value in the geosciences. 

In geoscientific applications, we often work with features $X $ that are functions of space, discretized in $N_x $ spatial locations $\boldsymbol{x} $, and of time, discretized in $N_t $ timesteps $\boldsymbol{t} $. We consider a model $M $ predicting a \textit{target} vector $\boldsymbol{Y} $ from a multi-variate spatiotemporal field $ \boldsymbol{X}_{\boldsymbol{x}, \boldsymbol{t}} $ such that:
\begin{equation}
    \boldsymbol{Y} = M\left[\boldsymbol{X}_{\boldsymbol{x}, \boldsymbol{t}}\right].
    \label{eq:model}
\end{equation}
We define a model $M $ as having added value w.r.t. a set of evaluation metrics $\left\{ {\cal E}_{i}\right\} _{i=1}^n $ and a baseline model $\widehat{M} $ if the following holds: When applied to representative out-of-sample data, $M $ is Pareto-optimal while the baseline model $\widehat{M} $ is not. Note that improving the Pareto front is necessary but not sufficient: a model predicting zero everywhere does not generally add value compared to other models solely because it is simpler.

\begin{figure*}[t]
  \centering
  \includegraphics[width=\textwidth]{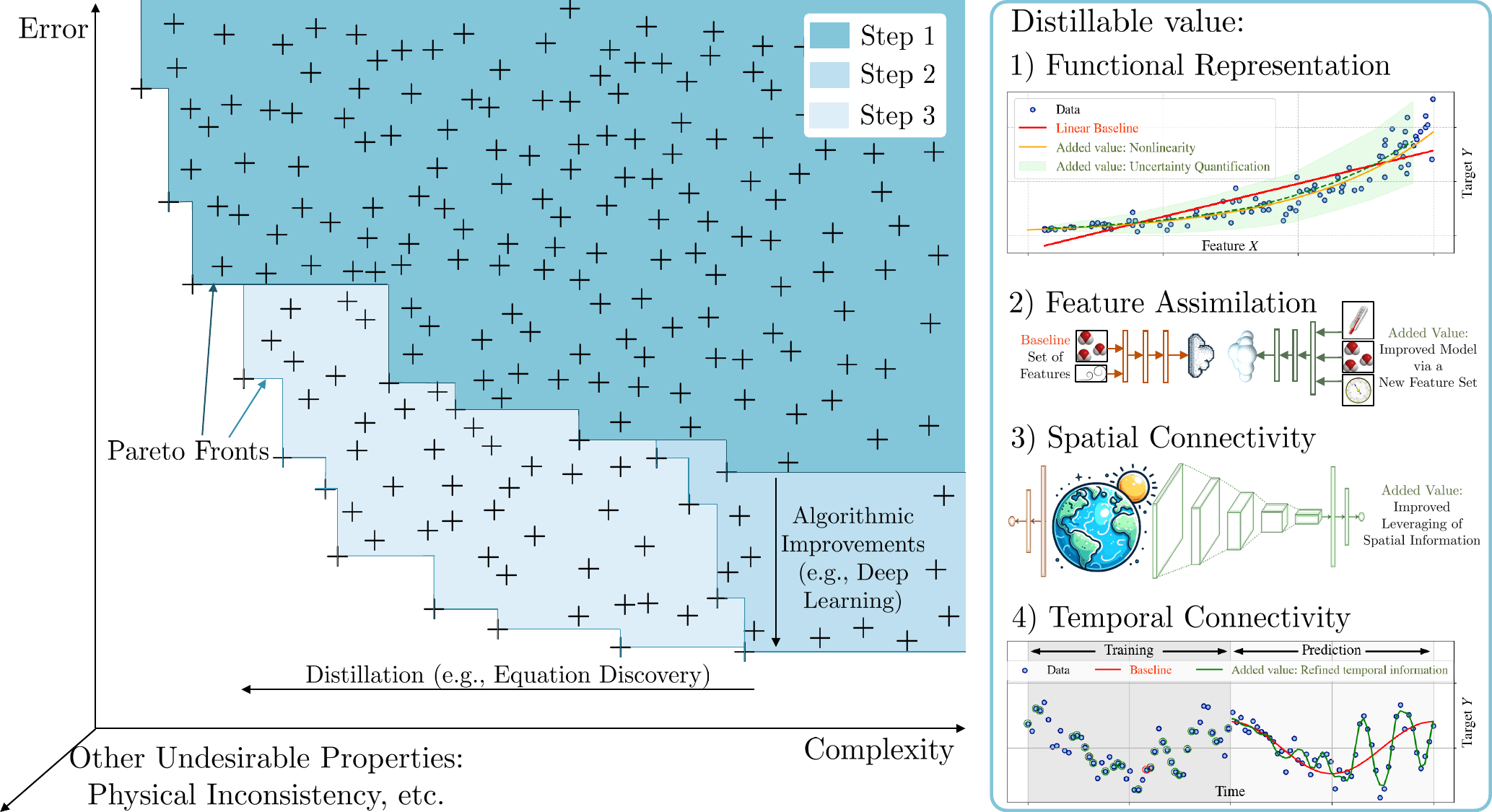}
  \caption{
Exploring Pareto fronts (sets of Pareto-optimal models) within a complexity-error plane highlights machine learning's added value. Crosses in step 1 denote existing models. Algorithms such as deep learning allow for the creation of efficient, low-error, albeit complex models (step 2). Knowledge distillation, through methods such as equation discovery, aims to explain error reduction, resulting in simpler, low-error models (step 3) and long-lasting scientific progress. For atmospheric applications, we propose four categories to classify this added value: functional representation, feature assimilation, spatial connectivity, and temporal connectivity.}\label{fig:pareto_opt_hierarchy}
\end{figure*}

Equation~\ref{eq:model} suggests four degrees of freedom for a model's added value: (1) improving the function $M $ without changing the features $\boldsymbol{X}_{\boldsymbol{x}, \boldsymbol{t}}$ (functional representation), (2) improving the model through $\boldsymbol{X} $ (feature assimilation), (3) improving the model through $\boldsymbol{x} $ (spatial connectivity), and (4) improving the model through $\boldsymbol{t} $ (temporal connectivity). We depict these four non-mutually-exclusive categories in Fig.~\ref{fig:pareto_opt_hierarchy} and rigorously define them below.

\subsection{Functional Representation}

A fundamental aspect a model can improve is the functional representation of the observed relationship between a set of baseline features $ \boldsymbol{\widehat{X}} $---\textit{fixed} features used to benchmark performance---and the target $ \boldsymbol{Y}$. We deem $M $ to improve the functional representation of a baseline $\widehat{M} $ w.r.t. ${\cal E} $ and ${\cal M} $ if and only if:
\begin{equation}
\widehat{M}\left[\widehat{\boldsymbol{X}}\right] \notin {\mathrm{PF}}_{{\cal E},{\cal M}}\ ,\  M\left[\widehat{\boldsymbol{X}}\right] \in {\mathrm{PF}}_{{\cal E},{\cal M}}
\label{eq:Fx_representation}
\end{equation}
This improvement may stem from algorithms leading to better fits (e.g., gradient boosting instead of decision trees), improved optimization (e.g., the Adam optimizer and its variants instead of traditional stochastic gradient descent; \citealp{kingma2014adam}), improved parsimony (e.g., by decreasing the number of trainable parameters via hyperparameter tuning) or enforced constraints (e.g., positive concentrations and precipitation). Improvements in functional representation are readily visualized via partial dependence plots or their variants (marginal plots to avoid unlikely data instances, accumulated local effects to account for feature correlation, etc.; \citealp{molnar2020interpretable}). Note that Eq.~\ref{eq:Fx_representation} also captures improvements in probabilistic modeling by generalizing $M $ to a stochastic mapping and including probabilistic scores \citep{gneiting2007strictly,haynes2023creating} in ${\cal E} $. From an atmospheric science perspective, improving functional representation helps identify nonlinear regimes, model extremes, and describe how sensitive the prediction is to different features. However, it can be challenging to faithfully describe these sensitivities if key features are missing from the baseline set of features $ \boldsymbol{\widehat{X}} $.

\subsection{Feature Assimilation}

This motivates discussing the ability of a model to extract relevant information from a new feature set, which we refer to as \textit{feature assimilation} and define as follows: 
\begin{equation}
\widehat{M}\left[\widehat{\boldsymbol{X}}\right] \notin {\mathrm{PF}}_{{\cal E},{\cal M}}\ ,\  M\left[\boldsymbol{X}\right] \in {\mathrm{PF}}_{{\cal E},{\cal M}}
\end{equation}
where we emphasize the difference between the baseline ($ \boldsymbol{\widehat{X}} $) and new ($ \boldsymbol{X} $) sets of features. This improvement may stem from previously unconsidered features (e.g., variables whose quality has increased in recent datasets), the ability of models to handle features whose information could not be extracted in past attempts (e.g., the ability of deep learning to extract nonlinear relationships), or the discovery of a more compact feature set that facilitates learning or improves generalizability. The assimilation of new features may improve interpretability (e.g, if some features simplify the target's functional representation or if there are less features to consider), physical consistency (e.g., if the new feature set improves generalizability across regimes or consistency with physical laws), and predictability. While the features' spatial and temporal discretizations could technically be considered part of feature selection, we choose to discuss them separately in the following subsections, given the central role of space and time in geoscientific applications. 

\subsection{Spatial connectivity}

To formalize a model $M $'s improved ability to leverage the features' spatial information, we distinguish the spatial locations $ \boldsymbol{\widehat{x}}$ used to discretize the baseline $\widehat{M} $'s features from the spatial locations $ \boldsymbol{x}$ used to discretize $M $'s features:
\begin{equation}
\widehat{M}\left[\widehat{\boldsymbol{X}}_{\widehat{\boldsymbol{x}},\widehat{\boldsymbol{t}}}\right] \notin {\mathrm{PF}}_{{\cal E},{\cal M}}\ ,\  M\left[\widehat{\boldsymbol{X}}_{\boldsymbol{x},\widehat{\boldsymbol{t}}}\right] \in {\mathrm{PF}}_{{\cal E},{\cal M}}
\end{equation}
This improvement may stem from the ability to: 
\begin{enumerate}
    \item handle features at different spatial resolutions (e.g., via improved pre-processing or handling of data). In atmospheric science, this can help consider multi-scale interaction and accommodate data from various Earth system models.
    \item hierarchically process spatially adjacent data (e.g., via convolutional layers). In atmospheric science, this acknowledges the high correlation between spatial neighbors due to, e.g., small-scale mixing. 
    \item capture long-range spatial dependencies (e.g., via self-attention mechanisms or a graph structure), such as teleconnections in the Earth system.
\end{enumerate}

\subsection{Temporal connectivity}

To formalize a model $M $'s improved ability to leverage the features' temporal information, we distinguish the timesteps $ \boldsymbol{\widehat{t}}$ used to discretize the baseline $\widehat{M} $'s features from the timesteps $ \boldsymbol{t}$ used to discretize $M $'s features:
\begin{equation}
\widehat{M}\left[\widehat{\boldsymbol{X}}_{\widehat{\boldsymbol{x}},\widehat{\boldsymbol{t}}}\right] \notin {\mathrm{PF}}_{{\cal E},{\cal M}}\ ,\  M\left[\widehat{\boldsymbol{X}}_{\widehat{\boldsymbol{x}},\boldsymbol{t}}\right] \in {\mathrm{PF}}_{{\cal E},{\cal M}}
\end{equation}
This improvement may stem from the ability to: 
\begin{enumerate}
    \item handle features at different temporal resolutions (e.g., via improved pre-processing or handling of data). This can help consider multiple timescales and accommodate data from various Earth system models.
    \item process consecutive timesteps (e.g., via recurrent layers). This acknowledges the high temporal autocorrelation of Earth system data, which is a property of the underlying dynamical system. 
    \item capture long-term temporal dependencies (e.g., via gating or self-attention mechanisms) and cyclic patterns (e.g., via data adjustments and temporal Fourier transforms) such as the diurnal and seasonal cycles.
\end{enumerate}

\section{Atmospheric Physics Application Cases \label{sec:Weather_Climate}}

This section demonstrates that Pareto optimality guides model development and improves process understanding through three realistic, atmospheric modeling case studies. Each case includes machine learning prototypes with demonstrated performance from previous studies, along with Pareto-optimal models newly trained for this study. The first case study emphasizes functional representation and feature assimilation, the second focuses on spatial connectivity, and the last compares spatial and temporal connectivity.



\subsection{Cloud Cover Parameterization for Climate Modeling\label{sub:application1}}



\subsubsection{Motivation}

The incorrect representation of cloud processes in current Earth system models, with grid spacing of approximately 50-100 km \citep{arias2021climate}, significantly contributes to structural uncertainty in long-term climate projections \citep{bony2015clouds,sherwood2014spread}. Cloud cover parameterization, which maps environmental conditions at the grid scale to the fraction of the grid cell occupied by clouds, directly affects radiative transfer and microphysical conversion rates, influencing the model's energy balance and water species concentrations.

Although ``storm-resolving'' simulations with grid spacing below $\approx5$ km do not explicitly resolve clouds and their associated microphysical processes \citep{morrison2020confronting}, they significantly reduce uncertainty in the interaction between storms and planetary-scale dynamics by explicitly simulating deep convection \citep{stevens2020added}. However, their large computational cost prohibits their routine use for ensemble projections \citep{schneider2017climate}. Machine learning can learn the storm-scale behavior of clouds from short, storm-resolving simulations, potentially improving coarser Earth system models through data-driven parameterizations \citep{gentine2021deep}.

\begin{figure*}[t]
  \centering
  \includegraphics[width=\textwidth]{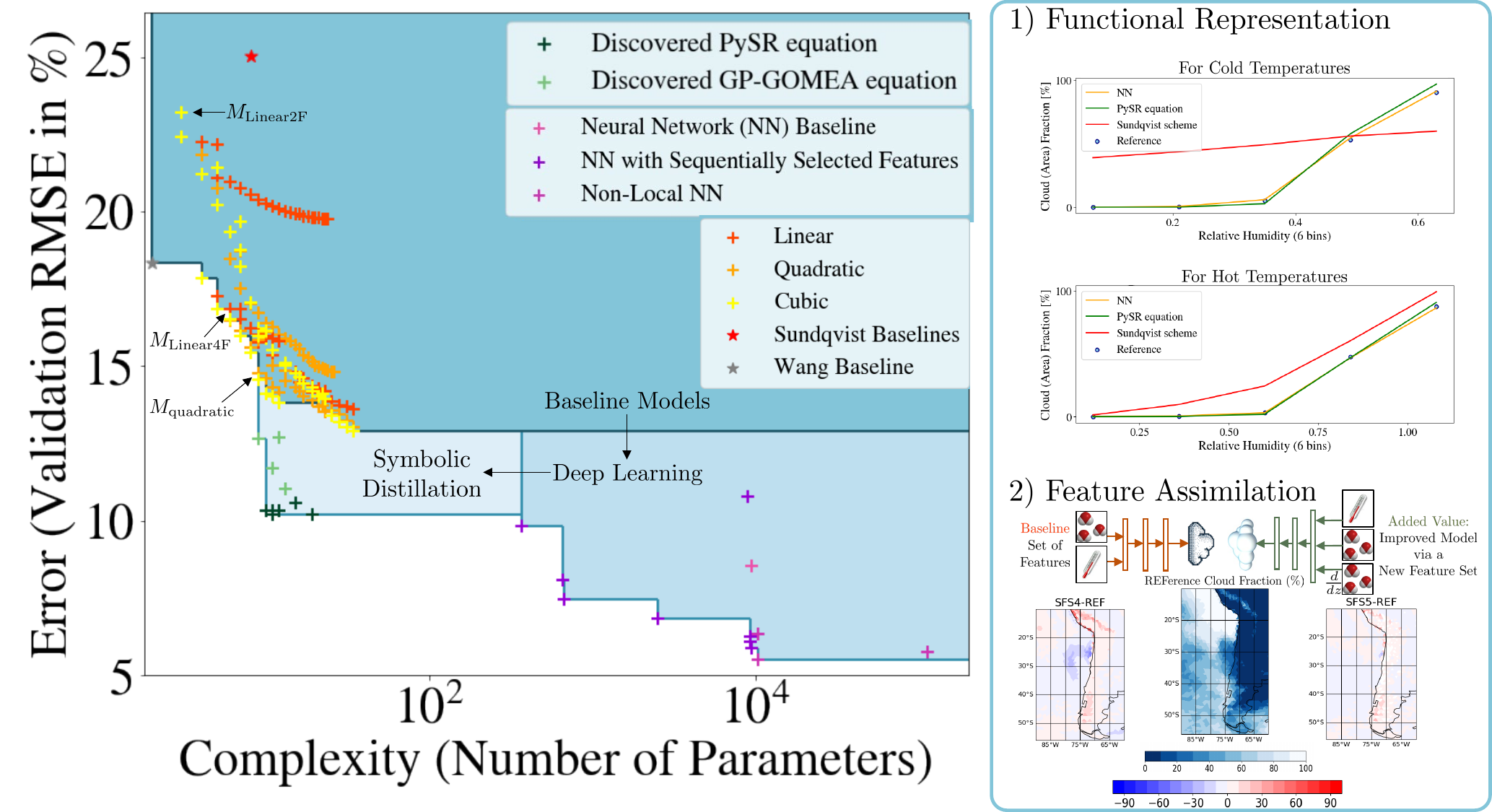}
  \caption{Pareto-optimal model hierarchies quantify the added value of machine learning for cloud cover parameterization. Machine learning better captures the relationship between cloud cover and its thermodynamic environment and assimilates features like vertical humidity gradients. (Left) We progressively improve traditional baselines via polynomial regression (red, orange, and yellow crosses), significantly decrease error using neural networks (pink and purple crosses), and finally distill the added value of these neural networks symbolically (green crosses). (Right) Both the neural network (orange line) and its distilled symbolic representation (green line) better represent the functional relationship between cloud cover and its environment, aligning more closely across temperatures with the reference storm-resolving simulation (blue dots) than the Sundqvist scheme (red line) used in the ICON Earth system model. ``Cold'' and ``Hot'' refer to the validation set's first and last temperature octiles. Additionally, machine learning models assimilate multiple features absent in existing baselines, including vertical humidity gradients. The smaller discrepancy between the 5-feature scheme (`SFS5') and the reference (`REF'), compared to the 4-feature scheme (`SFS4'), demonstrates improved representation of the time-averaged low cloud cover in regions such as the Southeast Pacific, thereby reducing biases in current cloud cover schemes that plague the global radiative budget.}\label{fig:cloud_cover}
\end{figure*}

This case study aims to understand the improvement gained from the higher-fidelity representation of storms and clouds. As illustrated in Fig.~\ref{fig:cloud_cover}, we demonstrate that this knowledge can be symbolically distilled into an analytic equation that rivals the performance of deep learning. 

\subsubsection{Setup\label{sub:setup}}

We follow the setup described in \cite{grundner2024data}, to which readers are referred for details. Fields are coarse-grained from storm-resolving ICON simulations \citep{giorgetta2018icon} conducted as part of the DYAMOND inter-comparison project \citep{stevens2019dyamond,duras2021dyamond}. The original simulations use a horizontal grid spacing of $\approx2.5$ km and 58 vertical layers below 21 km (the maximum altitude with clouds in the dataset). They are coarse-grained to a typical climate model resolution of $\approx80$ km horizontally and 27 vertical layers, converting the binarized, high-resolution condensate field (1 if the cloud condensate mixing ratio exceeds $10^{-6}$ kg/kg, 0 otherwise) into a fractional area cloud cover $\cal{C}$ (unitless).

To prevent strong correlations between the training and validation sets, the union of the ``DYAMOND Summer'' (Aug 10 to Sep 10, 2016) and ``DYAMOND Winter'' (Jan 30 to Feb 29, 2020) datasets was partitioned into six consecutive temporal segments. The second segment (approximately Aug 21 to Sep 1, 2016) and the fifth segment (approximately Feb 9–19, 2020) form the validation set. For all models, excluding traditional methods, the features are standardized to have a mean of zero and a standard deviation of one within the training set.

Once coarse-grained and pre-processed, we aim to map the environmental conditions $\boldsymbol{X}$ on vertical levels indexed by the background terrain-following height grid $\boldsymbol{z} $, to the cloud cover $\cal{C}$ on the same vertical levels. The variables $\boldsymbol{X}$ include the horizontal wind speed $\boldsymbol{U}$ [m/s], specific humidity $\boldsymbol{q_v}$ [kg/kg], liquid water mixing ratio $\boldsymbol{q_\ell}$ [kg/kg], ice mixing ratio $\boldsymbol{q_i}$ [kg/kg], temperature $\boldsymbol{T}$ [K], pressure $\boldsymbol{p}$ [Pa], and relative humidity $\boldsymbol{\mathrm{RH}}$. Except for the ``non-local NN'' in the next sub-section, we simplify the mapping to a ``vertically quasi-local'' one where cloud cover at a given level depends only on the atmospheric variables $\boldsymbol{X}$ at the same level and their first and second-order derivative with respect to $\boldsymbol{z}$. $\boldsymbol{X}$ also includes geometric height $z$ [m] and surface variables: land fraction $\sigma_f$ [\%] and surface pressure $p_s$ [Pa]. In summary, we approximate some mapping:
\begin{equation}
\underbrace{\left( \boldsymbol{X},\frac{d\boldsymbol{X}}{d\boldsymbol{z}}, \frac{d^2\boldsymbol{X}}{d\boldsymbol{z}^2}, z, \sigma_{f},p_{s} \right)}_{\in \mathbb{R}^{3\times7+3}} \mapsto\underbrace{\cal C}_{\in  \left[0, 1\right]}
\label{eq:quasi_local_mapping}
\end{equation}
using a hierarchy of machine learning models. In the following subsections, we will show that Pareto-optimal hierarchies not only facilitate data-driven model development, allowing for the comparison of simple baselines with neural networks, but also promote process understanding.


\subsubsection{Model Hierarchy\label{sub:NN}}

We start with the Sundqvist baseline (red star in Fig.~\ref{fig:cloud_cover}; \citealp{sundqvist1989condensation}), the standard cloud cover parameterization in ICON. The Sundqvist parameterization implemented in ICON represents cloud cover as a monotonically increasing function of relative humidity, provided it exceeds a critical threshold $\mathrm{RH}_\mathrm{crit}$ \citep{Roeckner1996TheAG}:
\begin{equation}
\mathrm{RH}_\mathrm{crit} \stackrel{\text{def}}{=} \mathrm{RH}_{\text{top}} + \left(\mathrm{RH}_{\text{surf}} - \mathrm{RH}_{\text{top}}\right) \exp \left(1 - \left[p_s / p\right]^n\right),
\end{equation}
where our implementation includes 4 tunable parameters: $\{\mathrm{RH}_{\text{surf}}, \mathrm{RH}_{\text{top}}, \mathrm{RH}_{\text{sat}}, n\}$, with values listed in appendix A.
When $\mathrm{RH} > \mathrm{RH}_\mathrm{crit}$, cloud cover is given by the model:
\begin{equation}
M_{\mathrm{Sundqvist}}:\ \left(p,p_s,\mathrm{RH}\right) \mapsto \mathcal{C}_{\mathrm{Sundqvist}},
\end{equation}
whose output is:
\begin{equation}
\mathcal{C}_{\mathrm{Sundqvist}} \stackrel{\text{def}}{=} 1 - \sqrt{\frac{\min \left\{\mathrm{RH}, \mathrm{RH}_{\text{sat}}\right\} - \mathrm{RH}_{\text{sat}}}{\mathrm{RH}_\mathrm{crit} - \mathrm{RH}_{\text{sat}}}}.
\label{eq:C_Sundqvist}
\end{equation}

To account for marine stratocumulus (low) clouds \citep{mauritsen2019developments}, we use two different sets of 4 parameters for land and sea using a land fraction threshold of 0.5. The Sundqvist scheme is parsimonious with only 8 trainable parameters, but it performs poorly against high-resolution data, with RMSE values as large as $25\%$ despite having been re-tuned to our training set.


Hypothesizing that the Sundqvist scheme's large error is due to its lack of features, we test the effect of adding features one-by-one. For that purpose, we apply forward sequential feature selection, which greedily adds features using a cross-validated score (here MSE), to a standard multiple linear regression model that includes polynomial combinations of all available features, up to a maximum degree of 3.

We find that linearly including temperature as a feature alongside relative humidity is enough to outperform the Sundqvist baseline, providing our simplest example of \textit{feature assimilation}:
\begin{equation}
M_{\mathrm{Sundqvist}}\left[\mathrm{RH}\right] \notin \mathrm{PF}_{{\cal E},{\cal M}}\ ,\  M_{\mathrm{Linear2F}}\left[\mathrm{RH}, T\right] \in \mathrm{PF}_{{\cal E},{\cal M}},
\label{eq:2F}
\end{equation}
where $M_{\mathrm{Linear2F}}$ is a two-feature linear regression whose output ${\cal C}_{\mathrm{Linear2F}} $ is:
\begin{equation}
\frac{{\cal C}_{\mathrm{Linear2F}}}{\alpha_{2F}} = 1 + \frac{\mathrm{RH}}{\mathrm{RH}_{2F}} - \frac{T}{\Delta T_{2F}}
\end{equation}
with parameters $\alpha_{2F}$, $\mathrm{RH}_{2F}$, and $\Delta T_{2F}$ listed in appendix A. This feature assimilation result corresponds to a well known result: accurate cloud cover parameterization requires temperature in addition to relative humidity.


Adding more features to linear (red crosses), quadratic (orange crosses), and cubic (yellow crosses) polynomials further reduces error. However, incorporating a simple physical constraint:
\begin{equation}
    q_{\ell}+q_{i}=0\ \Rightarrow\ {\cal C}=0,
\end{equation}
i.e., that the absence of condensates implies no clouds, further improves the results. Note that this constraint can also be derived from data using a binary decision tree. Accounting for this constraint, polynomial models achieve MSEs less than half of the Sundqvist scheme's (see Fig~\ref{fig:cloud_cover}'s lower error, higher complexity polynomial models). Focusing on the Pareto-optimal, four-feature linear and quadratic model outputs:
\begin{equation}
    \frac{{\cal C}_{\mathrm{Linear4F}}}{\alpha_{1}}=\ensuremath{1+\frac{\mathrm{RH}}{\mathrm{RH}_{1}}-\frac{T}{\Delta T_{1}}+\frac{q_{i}}{q_{1}}-H_{1}\frac{d\mathrm{RH}}{dz}},
    \label{eq:Multilinear}
\end{equation}
\begin{equation}
\begin{aligned}
\frac{{\cal C}_{\mathrm{quadratic}}}{\alpha_{2}} & =1+\frac{\mathrm{RH}}{\mathrm{RH}_{2,1}}-\frac{T}{\Delta T_{2,1}} + \frac{q_i}{q_{2,1}} + \frac{q_\ell}{q_{2,2}} -\frac{q_{i}q_{\ell}}{q_{2,3}^{2}}\\
 & +H_{2}\left(1+\frac{\mathrm{RH}}{\mathrm{RH}_{2,2}}-\frac{T}{\Delta T_{2,2}}\right)\frac{d\mathrm{RH}}{dz},
\end{aligned}
\label{eq:Quadratic}
\end{equation}
with parameters listed in appendix A. Analyzing this first Pareto front unveils the role of each added feature in better representing cloud cover. First, it highlights the role of cloud condensates, both liquid and ice, in accurately describing cloud cover. Second, among all possible vertical gradients, Pareto-optimal models select the negative gradient of relative humidity to depict the role of inversions in increasing cloud cover, which is particularly important for low clouds, including marine stratocumulus clouds. However, upon closer inspection, these contributions are not always physically consistent: while the four-feature linear model correctly captures the increase of cloud cover with ice, it incorrectly decreases cloud cover when there is humidity above. Conversely, the quadratic model accurately captures the inversion role when temperature and humidity are fixed to their training set mean but wrongly assumes that inversions' effect on cloudiness depends on humidity and temperature, and the model's decrease in cloudiness with increasing mixed-phase cloud condensates suggests that its $q_i q_\ell$ term is not an interpretable sensitivity to liquid and ice but rather a bias correction term relying on these variables.



Combining these insights with the Pareto-optimality of the simplified ``Wang'' scheme (gray star in Fig.~\ref{fig:cloud_cover}; \citealp{wang2023evaluating}, based on the scheme of \citealp{xu1996semiempirical}), which outputs:
\begin{equation}
    {\cal C}_{\mathrm{Wang}}=\mathrm{min}\left\{ 1,\left[1-e^{-\alpha_{\mathrm{Wang}}\left(q_{\ell}+q_{i}\right)}\right] \times \mathrm{RH}^{\beta_{\mathrm{Wang}}} \right\}, 
\end{equation}
with $\left(\alpha_{\mathrm{Wang}},\beta_{\mathrm{Wang}}\right) $ in appendix A, suggests that the dependence of cloud cover on condensates is fundamentally nonlinear. This justifies the use of deep learning to quickly explore which features can be combined to improve the nonlinear, \textit{functional representation} of cloud cover.



We start with the baseline neural network from \cite{grundner2022deep}: a 3-layer-deep, 64-neuron-wide multi-layer perceptron with batch normalization after the second hidden layer. Hyperparameters were optimized using the SHERPA Python library \citep{hertel2020sherpa}. This NN estimates the target cloud cover with high fidelity (RMSE=$8.5\%$), but at the cost of increased complexity, as it has a total of 9345 trainable parameters. Note that the models from \cite{grundner2022deep} were not designed to minimize the number of trainable parameters, so we do not overly focus on this complexity metric. The RMSE can be further lowered to $6\%$ with vertically non-local NNs, which map the entire atmospheric column of inputs to the entire column of outputs without inductive bias. However, this small error gain is deemed insufficient given the increased complexity cost.

Instead, we make the ``vertically quasi-local'' assumption (Eq.~\ref{eq:quasi_local_mapping}) and deploy a hierarchy of Pareto-optimal NNs, with features selected sequentially using cross-validated MSE. The five most informative features for NNs are:
\begin{equation}
\text{RH} \rightarrow q_i \rightarrow q_\ell \rightarrow T \rightarrow \frac{d\text{RH}}{dz}.     
\label{eq:SFS_list}
\end{equation}
Unlike features selected by polynomial models, these can be nonlinearly combined to yield high-quality predictions, as shown in the right panels of Fig.~\ref{fig:cloud_cover}. First, NNs improve the functional representation of cloud cover by accurately modeling the sharp increase in cloud cover above a temperature-dependent relative humidity threshold \textendash \, a highly nonlinear, bivariate behavior that simple schemes struggle to capture. Formally: 
\begin{equation}
    M_{\mathrm{Sundqvist}}\left[\boldsymbol{X}\right] \notin \mathrm{PF}_{{\cal E},{\cal M}}\ ,\  M_{\mathrm{NN}}\left[\boldsymbol{X}\right] \in \mathrm{PF}_{{\cal E},{\cal M}},
\end{equation}
where $M_{\mathrm{NN}} $ is a Pareto-optimal NN at the bottom-right of the (complexity, error) plane. Second, by incorporating vertical relative humidity gradients, NNs can capture stratocumulus decks off the coasts of regions like California, Peru/Chile, and Namibia/Angola. This improvement is especially visible when comparing the error map of an NN using the first four features from Equation~\ref{eq:SFS_list} to that of an NN additionally incorporating $d\text{RH}/{dz}$ (Fig~\ref{fig:cloud_cover}, bottom-right panel).

While indicative of how accurately cloud cover can be parameterized, such improvements are often insufficient to be considered ``discoveries'' as they remain hard to explain, even with post-hoc explanation tools (Fig.~8 \& 9 of \citealp{grundner2022deep}). Therefore, improvements in functional representation and feature assimilation need to be further distilled into sparse models that scientists can readily interpret.

\subsubsection{Symbolic Distillation and Equation Discovery\label{sub:eq_disc}}

For this purpose, we use symbolic regression libraries, which optimize both the parameters and structure of an analytical model within a space of expressions. Symbolic regression yields expressions with transparent out-of-distribution behavior (asymptotics, periodicity, etc.) \citep{la2021contemporary}, making them well-suited for high-stakes societal applications \citep{rudin2019stop} and the empirical distillation of natural laws \citep{schmidt2009distilling}. To avoid overly restricting the analytical form of the distilled equation, genetic programming is used. Genetic programming evolves a population of mathematical expressions using methods such as selection, crossover, and mutation to improve a fitness function \citep{koza1994genetic}. Given that genetic programming scales poorly with the number of features \citep{petersen2019deep}, our NN feature selection results are used to restrict our features to those listed in Equation \ref{eq:SFS_list}. Using NN results is appropriate since no assumption is made about the type of equation to be discovered.

The GP-GOMEA (light green crosses in Fig.~\ref{fig:cloud_cover}; \citealp{virgolin2021improving}) and PySR (dark green crosses in Fig.~\ref{fig:cloud_cover}; \citealp{cranmer2023interpretable}) libraries were chosen for their ease of use and high relative performance compared to 12 other recent libraries \citep{la2021contemporary}. They yielded over 500 closed-form equations for cloud cover, from which the 9 most physically consistent and lowest-error fits were retained with their outputs always clipped to $\left[0,1\right]$ (see \citealp{grundner2024data} for details). By physically interpreting the learned parameters, the output of the Pareto-optimal PySR model may be written as:
\begin{equation}
C_{\mathrm{PySR}}=\underbrace{{\cal I}_{1}\left(\mathrm{RH},T\right)}_{\mathrm{Humidity/Temperature}}+\underbrace{{\cal I}_{2}\left(\frac{d\mathrm{RH}}{dz}\right)}_{\mathrm{Inversion}}+\underbrace{{\cal I}_{3}\left(q_{\ell},q_{i}\right)}_{\mathrm{Condensates}},
\end{equation}
where the first term ${\cal{I}}_1$ may be interpreted as a sparse, third-order Taylor expansion around the training-mean relative humidity ($\overline{\mathrm{RH}}=0.60$) and temperature ($\overline{T}=257K$):
\begin{equation}
\begin{array}{cc}
{\cal I}_{1} & =\overline{{\cal C}}+\left(\frac{\partial{\cal C}}{\partial\mathrm{RH}}\right)_{\overline{\mathrm{RH}},\overline{T}}\left(\mathrm{RH}-\overline{\mathrm{RH}}\right)-\left(\frac{\partial{\cal C}}{\partial T}\right)_{\overline{\mathrm{RH}},\overline{T}}\left(T-\overline{T}\right)\\
 & +\frac{1}{2}\left(\frac{\partial^{2}{\cal C}}{\partial\mathrm{RH}^{2}}\right)_{\overline{\mathrm{RH}},\overline{T}}\left(\mathrm{RH}-\overline{\mathrm{RH}}\right)^{2}\\  
 & +\frac{1}{2}\left(\frac{\partial{\cal C}}{\partial\mathrm{RH}\partial T^{2}}\right)_{\overline{\mathrm{RH}},\overline{T}}\left(T-\overline{T}\right)^{2}\left(\mathrm{RH}-\overline{\mathrm{RH}}\right).
\end{array}
\label{eq:I1}
\end{equation}
Dominant Taylor series expansion terms are expected when discovering closed-form, subgrid-scale parameterizations \citep{jakhar2024learning}, but as PySR is based on genetic programming, we find two more surprising terms:
\begin{equation}
    {\cal I}_{2}=H_{\mathrm{PySR}}^{3}\left[\frac{d\mathrm{RH}}{dz}+\frac{3}{2}\left(\frac{d\mathrm{RH}}{dz}\right)_{\mathrm{max\ {\cal C}}}\right]\left(\frac{d\mathrm{RH}}{dz}\right)^{2},
    \label{eq:I2}
\end{equation}
where $H_{\mathrm{PySR}}\approx 585m$ can be interpreted as a characteristic height for low-cloud humidity gradients, and $\left(d\mathrm{RH}/{dz}\right)_{\mathrm{max\ {\cal C}}}$ ($\approx -2/\mathrm{km}$) is the value of the relative humidity gradient that maximizes cloud cover at the inversion level. The last term is:
\begin{equation}
    {\cal I}_{3}=-\frac{1}{\epsilon_{\mathrm{PySR}}} \times \frac{1}{1+2\epsilon_{\mathrm{PySR}}\left(\lambda_{\ell}q_{\ell}+\lambda_{i}q_{i}\right)},
    \label{eq:I3}
\end{equation}
which is a monotonically increasing function of the condensates' concentrations, whose trainable parameters are provided in appendix A. Consistently, the Pareto-optimal GP-GOMEA equation has a term ${\cal C}_{q}$ that sharply increases as liquid or ice concentrations exceed 0:
\begin{equation}
\frac{{\cal C}_{\mathrm{GOMEA}}}{\alpha_{G}}=1+\underbrace{\beta_{G}e^{\mathrm{RH}/\mathrm{RH_{G}}}}_{\mathrm{Humidity}}+\underbrace{{{\cal C}_{q}}\left(q_{\ell},q_{i}\right)}_{\mathrm{Condensates}},
\end{equation}
\begin{equation}
{\cal C}_{q}=\gamma_{G}\ln\left[\epsilon_{\mathrm{G}}+\frac{q_{i}}{q_{G,i}}+\delta_{G}\left(e^{\frac{q_{\ell}}{q_{G,\ell+}}}-1\right)\right]-\frac{q_{\ell}}{q_{G,\ell-}},
\label{eq:Q}
\end{equation}
where the trainable parameters are listed in appendix A. 

In addition to being easily transferable thanks to their low number of trainable parameters, the added value of these equations is transparent: the improved functional representation is explicit (see Eq.~\ref{eq:I1}), and the assimilation of new features is interpretable (see Eq.~\ref{eq:I2}). Finally, scientific discovery may arise through the unexpected aspects of these equations that are robust across models, such as the difference between how cloud cover reacts to an increase in environmental liquid versus ice content. Indeed, at high resolution, cloud cover will become 1 as soon as condensates exceed a small threshold (here $10^{-6}$ kg/kg) independently of the water's phase. Then, how come cloud cover is more sensitive to ice than liquid in Eq.~\ref{eq:I3} ($\lambda_i\approx 3.8\lambda_\ell $) and ${\cal C}_{q} $ increases much faster with ice than liquid (for $q_i+q_\ell\ll 1$) in Eq.~\ref{eq:Q}? 

These are in fact emerging properties of the subgrid distributions of liquid and ice \citep{grundner2024data}: As large values of cloud ice are rarely observed, larger spatial averages of cloud ice at coarse resolution means that many more high-resolution pixels contain low values of cloud ice compared to the liquid case, resulting in higher cloudiness for a given spatially-averaged condensate value. By assuming an exponential distribution for the subgrid liquid and ice content, we can even interpret $\lambda_\ell $ and $\lambda_i $ as the rate of the respective exponential distributions. This allows us to hypothesize that the nonlinear relationship between condensates and cloud cover is not scale-invariant and requires separate treatments of liquid and ice, with implications for the interaction between microphysical processes and the radiative budget. While analyzing the feature importance of liquid and ice in neural networks could have suggested this difference, it would have been difficult to fully explain it and bridge spatial scales without distillation, confirming the importance of only treating deep learning as a first and not final step towards knowledge discovery.

\subsection{Shortwave Radiative Transfer Emulation to Accelerate Numerical Weather Prediction\label{subsec:SW_Rad_Transfer}}

\subsubsection{Motivation}

The energy transfer resulting from the interaction between electromagnetic radiation and the atmosphere, known as radiative transfer, is costly to simulate accurately. Line-by-line calculations of gaseous absorption at each electromagnetic wavelength \citep{clough1992line} are too expensive for routine weather and climate models. Instead, models often use the correlated-$k$ method \citep{mlawer1997radiative}, which groups absorption coefficients in a cumulative probability space to speed up radiative transfer calculations without significantly compromising accuracy. However, even the correlated-$k$ method imposes a high computational burden \citep{veerman2021predicting}, forcing most simulations to reduce the temporal and spatial resolution of radiative transfer calculations, which can degrade prediction quality \citep{morcrette2000effects,hogan2018flexible}.

This challenge has driven the development of ML emulators for radiative transfer in numerical weather prediction since the 1990s \citep{cheruy1996methode,chevallier1998neural,chevallier2000use}. ML architectures have become more sophisticated (e.g., \citealp{belochitski2021robustness,kim2022usefulness,ukkonen2022exploring}), but the primary goal remains to emulate the original radiation scheme as faithfully as possible. This allows the reduced inference cost of the ML model, once trained, to be leveraged for running the atmospheric model coupled with the emulator, enabling less expensive and more frequent radiative transfer calculations.

\begin{figure*}[t]
  \centering
  \includegraphics[width=\textwidth]{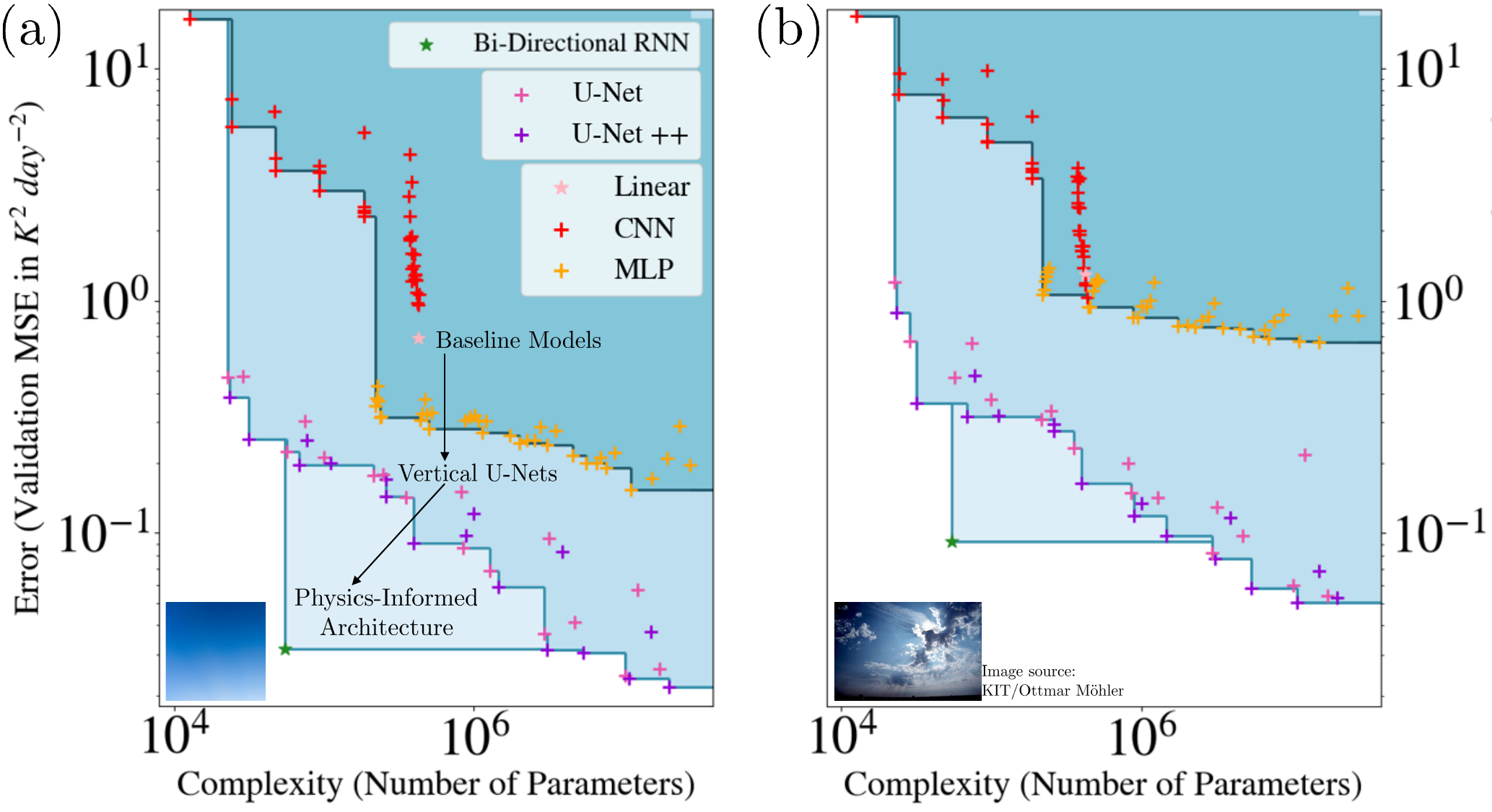}
  \caption{Pareto-optimal model hierarchies guide the development of progressively tailored architectures for emulating shortwave radiative transfer. Panel (a) shows error vs. complexity on a logarithmic scale for the simple clear-sky cases dominated by absorption; panel (b) shows error vs. complexity for cases with multi-layer cloud, including both liquid and ice, where multiple scattering complicates radiative transfer. Convolutional neural networks (CNN; red crosses) with small kernels, multilayer perceptrons (MLP; orange crosses) that ignore the vertical dimension, and the simple linear baseline (light pink star) give credible results in the clear-sky case. However, they fail in the more complex case, which requires U-net architectures (dark pink and purple crosses) to fully capture non-local radiative transfer. The vertical invariance of the two-stream radiative transfer equations suggests a bidirectional recurrent neural network (RNN; green star) architecture, which rivals the skill of U-nets with a fraction of their trainable parameters.}\label{fig:sw_rad}
\end{figure*}

This case study examines how ML architectural designs impact the reliability of shortwave radiative transfer (covering solar radiation and wavelengths of 0.23-5.85 $\mu$m). As shown in Fig.~\ref{fig:sw_rad}, physics-informed architectures that closely mimic the vertical bidirectionality of radiative transfer are Pareto-optimal, rivaling the performance of deep learning models with ten times more trainable parameters.

\subsubsection{Setup}

We follow the setup described in \cite{lagerquist2023estimating} and emulate the full behavior (gas optics and radiative transfer solver) of the shortwave Rapid Radiative Transfer Model \citep{mlawer1997radiative} in the context of weather predictions made by version 16 of the Global Forecast System with $0.25^{\circ}$ horizontal spacing. In addition to the realistic geography setup of \cite{lagerquist2021using}, the ML models are trained with global data on the 127-level native pressure-sigma grid (henceforth referred to as $\boldsymbol{\eta}$) with synthetic information about aerosols, trace gases, and hydrometeors' particle size distribution. We use data from most days between Sep 1 2018 and Dec 23 2020, with forecast lead times in $\left\{6, 12, 18, 24, 30, 36\right\}$ hours. In each forecast, 4000 grid points are randomly sampled on the global grid. The training set comprises 873,086 samples from 237 days, the validation set 479,806 samples from 126 days, and the test set 472,456 samples from 120 days. Features $\boldsymbol{X}$ can broadly be separated into 22 vertical profiles tied to the grid ($\boldsymbol{X_\eta}$; listed in Appendix B) and four scalars with one value per atmospheric column: solar zenith angle $\zeta$ [$^{\circ}$], surface albedo $\alpha_s$, aerosol single-scattering albedo $\omega_0$, and aerosol asymmetry parameter $g$. For convenience, the four scalars are converted into profiles by repeating their values at all 127 levels. We target the shortwave heating rate's vertical profile $\left(\boldsymbol{dT}/\boldsymbol{dt}\right)_{\text{SW}}$ [K day\textsuperscript{-1}], thus approximating a high-dimensional mapping:
\begin{equation}
\underbrace{\left(\boldsymbol{X_{\eta}},\zeta,\alpha_{s},\omega_{0},g,\sigma_{e}\right)}_{\in\mathbb{R}^{\left(22+4\right)\times127}} \mapsto \underbrace{\boldsymbol{\left(\frac{dT}{dt}\right)_{\text{SW}}}}_{\in\mathbb{R}^{1\times127}}
\end{equation}
via an ML model hierarchy described in the next section.

\subsubsection{Model Hierarchy}

We deploy an ML model hierarchy using the same features $\boldsymbol{X}$ to isolate the added value of spatial connectivity along the vertical coordinate $\boldsymbol{\eta}$. For all models, cropping and zero-padding layers set the top-of-atmosphere (TOA; $\sim$78 km above ground level) heating rate to zero, consistent with the RRTM model being emulated. Readers interested in the exact ML model architecture are referred to \cite{lagerquist2023estimating}.

We begin with a linear regression model (pink star), where the input layer is reshaped into a flattened vector and processed through linear layers without activation functions or pooling. This model lacks inherent spatial connectivity, treating all features (where one ``feature'' = one atmospheric variable at one vertical level) independently. Similarly, the multilayer perceptrons (MLP; orange crosses) flatten inputs, ignoring vertical connectivity, but process them with varying degrees of nonlinearity, controlled by two hyperparameters: depth (number of dense layers) and width (number of neurons per layer). We conduct a grid search with depth varying from $\lbrace1, 2, 3, 4, 5, 6\rbrace$ layers and width varying from $\lbrace64, 128, 256, 512, 1024, 2048\rbrace$ neurons, resulting in 36 MLPs. The activation function for every hidden layer (\textit{i.e.}, every layer except the output) is the leaky rectified linear unit \citep[ReLU;][]{Maas2013} with a negative slope $\alpha = 0.2$.

To incorporate vertical relationships between different levels, we train convolutional neural networks (CNN; red crosses) with 1D convolutions along the vertical axis. We conduct a grid search with depth varying from $\lbrace1, 2, 3, 4, 5, 6\rbrace$ convolutional layers and number of channels in the first convolutional layer varying from $\lbrace2, 4, 8, 16, 32, 64\rbrace$. After the first layer, we double the number of channels in each successive convolutional layer, up to a maximum of 64. The CNN's kernel size is restricted to 3 vertical levels, to enforce strong local connectivity.

The U-net architecture (pink crosses; \citealp{ronneberger2015u}) builds on the CNN by incorporating skip connections that preserve high-resolution spatial information and an expansive path almost symmetric to the contracting path. Both of these model components improve the reconstruction of full-resolution data (here, a 127-level profile of heating rates). We conduct a grid search with depth (number of downsampling operations) varying from $\lbrace3, 4, 5\rbrace$ and number of channels in the first convolutional layer varying from $\lbrace2, 4, 8, 16, 32, 64\rbrace$, using the same channel-doubling rule as for CNNs. Finally, the U-net++ architecture (purple crosses; \citealp{zhou2019unet++}) enhances the U-net's skip pathways with nested dense convolutional blocks, potentially capturing more of the original spatial information and facilitating optimization. Our hyperparameters for U-net++ are the same as for the U-net.

\subsubsection{Distilling Radiative Transfer's Bidirectionality}

To better understand the added value of each architecture in representing shortwave absorption and scattering, we extract samples from our test set to form two distinct regimes. The simple, ``clear-sky'' regime (Fig.~\ref{fig:sw_rad}a) consists of profiles with no cloud (column-integrated liquid-water path = column-integrated ice-water path = 0 g m\textsuperscript{-2}), an oblique Sun angle (zenith angle $>$ 60$^{\circ}$), and little water vapor (column-maximum specific humidity $<$ 2 g kg\textsuperscript{-1}). This restricts shortwave radiative transfer to gaseous absorption of solar radiation throughout the atmospheric column, well described by a simple exponential attenuation model such as Beer's law \citep{liou2002introduction}. In contrast, the complex, ``multi-cloud'' regime (Fig.~\ref{fig:sw_rad}b) includes profiles with more than one cloud layer and at least one mixed-phase cloud layer.  For this purpose, a ``cloud layer'' is defined as a set of consecutive vertical levels, such that every level has a total water content (liquid + ice) $>$ 0 g m\textsuperscript{-3} and the layer has a height-integrated total water path (liquid + ice) $\ge$ 10 g m\textsuperscript{-2}.  A mixed-phase cloud layer meets the above criteria plus two additional ones: both the height-integrated liquid water path and ice water path must be $>$ 0 g m\textsuperscript{-2}. This regime is challenging to model as shortwave radiation is absorbed and scattered by both liquid and ice clouds, making the bidirectionality of radiative transfer essential to capture. Out of the 472,456 test set profiles, 14,226 are in the clear-sky regime and 13,263 in the multi-cloud regime.

The first surprising result is the poor performance of CNNs: While Pareto-optimal for low complexity, with our simplest CNN having as few as 12,630 trainable parameters, none of the CNNs achieve an MSE below 0.95 K\textsuperscript{2} day\textsuperscript{-2}, even in the clear-sky case. In contrast, MLPs systematically outperform our linear baseline model, showcasing the importance of nonlinearity. MLPs also outperform CNNs because they allow for non-local vertical connectivity. However, without information on which vertical levels are closest, MLPs connect every level together, resulting in high complexity: Increasing the number of trainable parameters by 100$\times$ does not even halve the MSE in the clear-sky case. This lack of inductive bias prevents generalization to more complex cases such as the multi-cloud regime, where the MSE climbs from $\approx$0.2-0.4 K\textsuperscript{2} day\textsuperscript{-2} to $\approx$0.7-1.0 K\textsuperscript{2} day\textsuperscript{-2}.

By ordering the features' vertical levels, both the U-net and U-net++ achieve lower errors with fewer trainable parameters. Most Pareto-optimal models are in the U-net++ family, suggesting that simple skip connections are insufficient; this highlights the non-locality of shortwave radiative transfer, especially in the multi-cloud case. Fortunately, while shortwave radiation may instantaneously propagate information (e.g., the presence of mixed-phase clouds) throughout the entire atmospheric column, its transfer is approximately governed by bidirectional equations invariant in space. The two-stream equation holds exactly for the target Rapid Radiative Transfer Model data:
\begin{equation}
    \frac{d}{d\eta}\begin{pmatrix}{\cal F}_{+}\\
    {\cal F}_{-}
    \end{pmatrix}={\cal R}\left({\cal F}_{+},{\cal F}_{-},X_\eta\right),
    \label{eq:2SRT}
\end{equation}
where ${\cal F}_{+}$ and ${\cal F}_{-}$ are the upward and downward radiative fluxes (W m\textsuperscript{-2}) and the function $\cal R$ governs how these fluxes change with the vertical coordinate $\eta$, allowing us to update fluxes using neighboring fluxes only. As noted in \cite{ukkonen2024representing}, it is the vertical invariance of Eq.~\ref{eq:2SRT} that suggests we may treat radiative transfer with a network processing the vertical sequences of ${\cal F}_{+}$ and ${\cal F}_{-}$ in both directions with the same update rule at every level. Akin to standard atmospheric models, the target heating rate is then calculated by multiplying the vertical gradient of the net flux (downward minus upward) by the ratio of the gravity constant at the Earth's surface $g$ to the specific heat capacity of dry air at constant pressure $c_p$:
\begin{equation}
    \left(\frac{dT}{dt}\right)_{\mathrm{SW}}=\frac{g}{c_p}\frac{d\eta}{dp}\frac{d\left({\cal F}_{-}-{\cal F}_{+}\right)}{d\eta}.
\end{equation}
Following \cite{ukkonen2022exploring}, we divide the shortwave radiative fluxes $\left({\cal F}_{+},{\cal F}_{-}\right)$ by the incoming TOA downward flux (${\cal F}_{-}^{\textrm{TOA}}$), to implicitly inform the NN that the Sun is the original energy source. At each vertical level $\eta$, we then target the resulting ratios (${{\cal F}_{+}}/{\cal F}_{-}^{\textrm{TOA}}$ and ${{\cal F}_{-}}/{{\cal F}_{-}}^{\textrm{TOA}}$) with a bidirectional RNN. The RNN has two 64-unit gated recurrent unit layers (one forward, one backward), which are concatenated before a dense layer for predicting the fluxes, resulting in only 54,786 trainable parameters. The RNN includes a multiplication layer with ${\cal F}_{-}^{\textrm{TOA}}$ and a custom layer to directly calculate shortwave heating rates, maintaining the same optimization objective as the other models. We find that the resulting RNN (green star) is clearly Pareto-optimal, competing with the U-net++ and achieving MSE below 0.1 K\textsuperscript{2} day\textsuperscript{-2}, even in the complex, multi-cloud case. Our results highlight the advantages of physically constraining ML solutions using robust knowledge of the underlying system, extending the aquaplanet findings of \cite{Bertoli_2023_partI} to an Earth-like, operational setting, and paving the way towards fully hybrid physics-ML emulators of shortwave radiation \citep{Schneiderman2024}. 

\subsection{Tropical Precipitation and Convective Organization}

\subsubsection{Motivation}

Accurately representing precipitation processes in tropical regions enhances global Earth system models and forecasting tools, particularly for water management and flood risk in a changing climate \citep{seneviratne2021weather}. Due to computational limitations, achieving horizontal resolutions below 25-50 km is challenging, which hinders the representation of subgrid processes causing precipitation \citep{stevens2020added}. These processes include tropical convection and its complex organization patterns \citep{bao2024intensification}, typically modeled using semi-empirical parameterizations that rely on a coarse representation of the thermodynamic environment and often exclude memory effects \citep{colin2019identifying}. These parameterizations generally approximate a mapping: $\left\langle \boldsymbol{X}_{t_{0}}\right\rangle\mapsto\left\langle P_{t_{0}}\right\rangle$, where at the time of interest $t_{0}$, both the spatially resolved environmental predictors $\boldsymbol{X}$ and the precipitation are averaged over the coarse grid cell:
\begin{equation}
    \left\langle \boldsymbol{X}_{t_{0}}\right\rangle=\frac{1}{\left| \mathrm{Grid\ Cell} \right|}\int_{\mathrm{Grid\ Cell}}\boldsymbol{X}_{t_{0},\boldsymbol{x}}d\boldsymbol{x},
\end{equation}
where $\boldsymbol{x}$ represents a continuous, bidimensional horizontal coordinate. This spatial averaging removes convective organization information below the coarse grid's spatial scale. Using storm-resolving simulations as a reference, \cite{shamekh2023implicit} showed this blurs precipitation, adding irreducible uncertainty, especially for large values of precipitable water ($Q $). Given that high-resolution information is typically inaccessible in a parameterization context, we ask in this case study: How much of this lost spatial granularity can we recover with temporal memory?

\subsubsection{Setup}

To address this question, we build upon the framework of \cite{shamekh2023implicit}, coarse-graining SAM-DYAMOND \citep{khairoutdinov2022global} simulations from their native, high-resolution horizontal grid (4.34-km spacing at the equator) to a low-resolution grid (138.9-km spacing at the equator) representative of a coarse Earth system model. Over the tropical ocean ($20^{\circ}$S-$20^{\circ}$N), we map 6 environmental variables $\boldsymbol{X}$ \textendash \, precipitable water [mm], 2-m specific humidity [kg kg\textsuperscript{-1}], 2-m temperature [K], surface sensible and latent heat fluxes [$\text{W}/\text{m}^2$], and sea surface temperature [K] \textendash \, to 15-min precipitation rates $\left\langle P\right\rangle$ [mm h\textsuperscript{-1}]. We use three setups at a given low-resolution location $x_{\mathrm{LR}}$ including the high-resolution grid locations $\boldsymbol{x_{\mathrm{HR}}}$:
\begin{equation}
    \text{Baseline: } \underbrace{\left\langle\boldsymbol{X}_{t_{0}}\right\rangle_{x_{\mathrm{LR}}}}_{\in \mathbb{R}^6}  \mapsto \underbrace{\left\langle P_{t_{0}}\right\rangle_{x_{\mathrm{LR}}}}_{\in \mathbb{R}_+},
    \label{eq:Baseline}
\end{equation}
\begin{equation}
    \text{Spatial Granularity: } \underbrace{\left(\begin{array}{c}
\left\langle\boldsymbol{X}_{t_0}\right\rangle_{x_{\mathrm{LR}}}\\
Q_{t_0,\boldsymbol{x_{\mathrm{HR}}}}^{\prime}
\end{array}\right)}_{\in \mathbb{R}^{6+32\times32}}\mapsto\underbrace{\left\langle P_{t_{0}}\right\rangle_{x_{\mathrm{LR}}}}_{\in \mathbb{R}_{+}},
    \label{eq:Spatial_Granularity}
\end{equation}
\begin{equation}
    \text{Temporal Memory: } \underbrace{\left\langle \boldsymbol{X}_{\boldsymbol{t_{\mathrm{Memory}}}}\right\rangle_{x_{\mathrm{LR}}}}_{\in\mathbb{R}^{6\times\text{card}\left(\boldsymbol{t_{\mathrm{Memory}}}\right)}}\mapsto\underbrace{\left\langle P_{t_{0}}\right\rangle_{x_{\mathrm{LR}}}}_{\in \mathbb{R}_{+}},
    \label{eq:Temporal_Memory}
\end{equation}
where $Q^{\prime}$ is the precipitable water's anomaly with respect to its grid cell-mean $\langle Q \rangle$ and memory is accounted for through up to two time steps in the past, in which case $\boldsymbol{t_{\mathrm{Memory}}}=\left\{ t_{0},t_{0}-15\text{min},t_{0}-30\text{min}\right\}$. Samples are extracted from 10 days of the simulation after spin-up, with 6 days for training, 2 for validation, and 2 for testing. To focus on precipitation intensity rather than triggering, samples with precipitation values below 0.05 mm/h are discarded, resulting in a total of $\approx 10^8$ samples.

\subsubsection{Model Hierarchy}
\begin{figure*}[t]
  \centering
  \includegraphics[width=\textwidth]{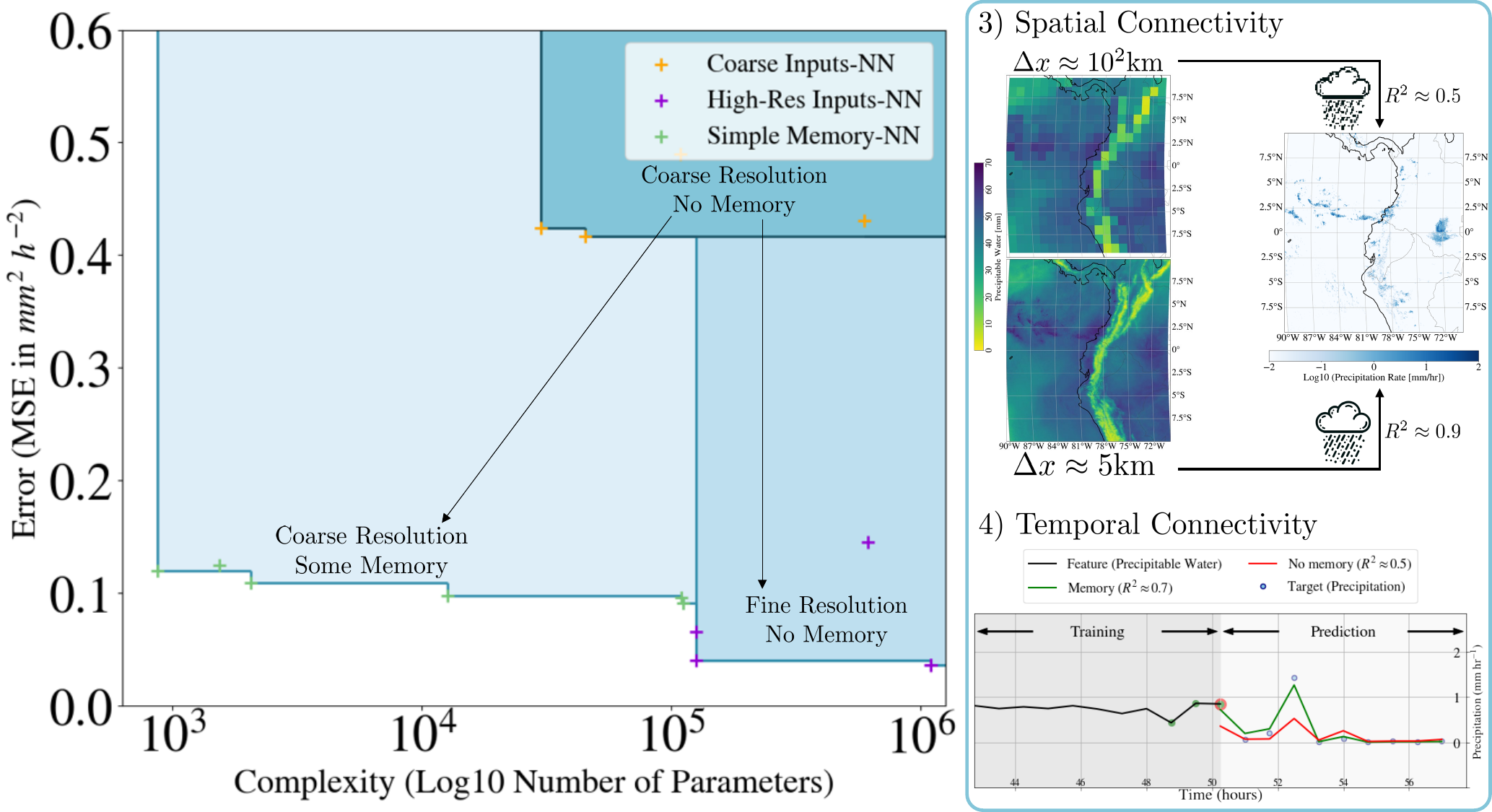}
  \caption{Pareto-optimal model hierarchies underscore the importance of storm-resolving information in elucidating the relationship between precipitation and its surrounding environment, while also quantifying the recoverability of this information from the coarse environment's time series. (Left) Neural networks (NN) leveraging high-resolution spatial data (purple crosses) clearly outperform NNs that use only coarse inputs (orange crosses). However, this performance gap is largely mitigated when the coarse inputs' past time steps are included (green crosses). (Right) Processing the precipitable water field at a resolution of $\Delta x \approx 5$ km yields coefficients of determination $R^2 \approx 0.9$, clearly surpassing the $R^2 \approx 0.5$ attained by our best NN using fields at the coarse $\Delta x \approx 10^2$ km horizontal resolution. This performance gap is partially closed by incorporating two past time steps along with the current timestep, resulting in $R^2 \approx 0.7$. This suggests a partial equivalence of the environment's spatial and temporal connectivities in predicting precipitation.}\label{fig:tropical_precipitation}
\end{figure*}

As illustrated in Fig.~\ref{fig:tropical_precipitation}, we design three categories of NNs to learn the three mappings given by Eq.~\ref{eq:Baseline}, \ref{eq:Spatial_Granularity}, and \ref{eq:Temporal_Memory}. First, our baseline NNs (orange crosses) are simple MLPs with 5 layers of different widths, resulting in a number of trainable parameters ranging from $\approx 30-600 \times 10^3$ and never achieving MSEs below 0.4 mm\textsuperscript{2} h\textsuperscript{-2}. The NNs using spatial granularity (purple crosses) additionally encode the high-resolution precipitable water anomaly field through an encoder-decoder structure with a small bottleneck representing convective organization as a bidimensional latent variable, optionally regularized via data augmentation applied to $Q^{\prime}$. MSEs are below 0.04 mm\textsuperscript{2} h\textsuperscript{-2} for the deepest encoder-decoder architectures, which yield high reconstruction quality through more than 1M trainable parameters. The NNs leveraging temporal memory instead of spatial resolution (green crosses) use the last or last two previous timesteps to achieve competitive MSEs, all below 0.15 mm\textsuperscript{2} h\textsuperscript{-2}. These ``Memory-NNs'' are built using two types of architectures: On the one hand, we flatten previous timesteps and feed them to MLPs conditioned on non-zero current precipitation, which ignores temporal connectivity. On the other hand, we use CNN-based temporal models that preserve temporal ordering through 1D convolutional layers, achieving MSEs around 0.1 mm\textsuperscript{2} h\textsuperscript{-2} with $\approx 10$ times fewer trainable parameters than the ``High-Res Inputs-NNs'' that use storm-scale information. Overall, our results showcase the added value of \textit{spatial and temporal connectivities}, as $M_{\mathrm{MLP}}\left[\left\langle\boldsymbol{X}_{t_{0}}\right\rangle_{x_{\mathrm{LR}}}\right]\notin\mathrm{PF}_{{\cal E},{\cal M}} $ while:
\begin{equation}
    M_{\mathrm{CNN}}\left[\left\langle \boldsymbol{X}_{\boldsymbol{t_{\mathrm{Memory}}}}\right \rangle_{x_{\mathrm{LR}}}\right]\in\mathrm{PF}_{{\cal E},{\cal M}},
\end{equation}

\begin{equation}
M_{\mathrm{ED}}\left[\left(\begin{array}{c}
\left\langle\boldsymbol{X}_{t_0}\right\rangle_{x_{\mathrm{LR}}}\\
Q_{t_0,\boldsymbol{x_{\mathrm{HR}}}}^{\prime}
\end{array}\right)\right]\in\mathrm{PF}_{{\cal E},{\cal M}}.    
\end{equation}
    
\subsubsection{Mitigating Low Spatial Resolution with Memory}

While it may be unsurprising that leveraging spatio-temporal information decreases error, the competitive MSEs obtained with temporal memory but without high-resolution spatial information are promising for improving parameterizations. We ask: How can we explain that temporal memory helps recover a large portion of the spatial granularity information?

First, we can ask: Why would temporal memory of the coarse environment $\left\langle \boldsymbol{X}_{\boldsymbol{t_{\mathrm{Memory}}}}\right\rangle_{x_{\mathrm{LR}}} $ be relevant on its own, given that at a given time $t_0$, the precipitation $\left\langle P_{t_{0}}\right\rangle $ can be exactly diagnosed from the high-resolution environment $\boldsymbol{X}_{t_{0},\boldsymbol{x_{\mathrm{HR}}}} $? The informativeness of high spatial resolution is confirmed by the fact that our lowest-error model is $M_{\mathrm{ED}} $. As written in Eq.~\ref{eq:Spatial_Granularity}, $M_{\mathrm{ED}} $ decomposes the high-resolution information into a low-resolution component and the inputs' spatial anomaly (denoted by primes). Using this decomposition, we can write the following causal graph:
\begin{equation}
    \begin{array}{c}
    
     \left\langle \boldsymbol{X}_{t_{0}-\Delta t}\right\rangle_{x_{\mathrm{LR}}}\rightarrow \left\langle \boldsymbol{X}_{t_{0}}\right\rangle_{x_{\mathrm{LR}}} \leftarrow \boldsymbol{X}^{\prime}_{t_{0}-\Delta t,\boldsymbol{x_{\mathrm{HR}}}},\\  

     \left\langle \boldsymbol{X}_{t_{0}-\Delta t}\right\rangle_{x_{\mathrm{LR}}} \textcolor{blue}{\rightarrow} \boldsymbol{X}^{\prime}_{t_{0},\boldsymbol{x_{\mathrm{HR}}}} \leftarrow \boldsymbol{X}^{\prime}_{t_{0}-\Delta t,\boldsymbol{x_{\mathrm{HR}}}},\\  
    \boldsymbol{X}^{\prime}_{t_{0},\boldsymbol{x_{\mathrm{HR}}}} \textcolor{blue}{\rightarrow} P_{t_{0}} \leftarrow \left\langle \boldsymbol{X}_{t_{0}}\right\rangle_{x_{\mathrm{LR}}}, 
    
    \end{array}
\label{eq:Causal_graph}
\end{equation}
where we posit that for a small enough time step $\Delta t$, the combination of the high-resolution anomaly and low-resolution information suffices to diagnose precipitation and step the entire system forward in time from $t_{0}-\Delta t $ to $t_{0} $. Formally, this is a Markovian assumption, which eliminates the need for temporal memory. Under this assumption, the precipitation $P_{t_{0}} $ is independent from $\left \langle \boldsymbol{X}_{t_{0}-\Delta t}\right\rangle_{x_{\mathrm{LR}}} $ and previous timesteps when conditioned on the current environment $\left( \boldsymbol{X}^{\prime}_{t_{0},\boldsymbol{x_{\mathrm{HR}}}},\left\langle \boldsymbol{X}_{t_{0}}\right\rangle_{x_{\mathrm{LR}}} \right)$. Using the bottom two lines of Eq.~\ref{eq:Causal_graph}, we see that suppressing access to high-resolution information activates the following causal path from $\left \langle \boldsymbol{X}_{t_{0}-\Delta t}\right\rangle_{x_{\mathrm{LR}}} $ to $P_{t_{0}} $ by introducing a dependency:
\begin{equation}
    \left \langle \boldsymbol{X}_{t_{0}-\Delta t}\right\rangle_{x_{\mathrm{LR}}} \textcolor{blue}{\rightarrow} \boldsymbol{X}^{\prime}_{t_{0},\boldsymbol{x_{\mathrm{HR}}}} \textcolor{blue}{\rightarrow} P_{t_{0}}.
\end{equation}
This induces a conditional dependence of precipitation $P_{t_{0}}$ on the past low-resolution environment $\left\langle \boldsymbol{X}_{t_{0}-\Delta t}\right\rangle_{x_{\mathrm{LR}}}$ by removing d-separation through the exclusion of current high-resolution anomalies $\boldsymbol{X}^{\prime}_{t_{0},\boldsymbol{x_{\mathrm{HR}}}}$ from the conditioning set \citep{pearl2000models}. Therefore, temporal memory becomes necessary in the absence of high-resolution information, consistent with the general result of the Mori-Zwanzig decomposition for dynamical systems \citep{lucarini2023theoretical}. This explains the Pareto optimality of the "simple-memory NN" $M_{\mathrm{CNN}}$ that considers low-resolution inputs and their past timesteps.

Now, we can ask: How much information is lost in the process of replacing the high-resolution spatial anomaly with temporal memory of the low-resolution spatial means? The answer depends on the distance between low and high resolutions. There is no difference if these resolutions are equal, while we do not expect low-resolution memory to help at all if the low-resolution grid is coarse to the point that we cannot distinguish very different high-resolution situations. For the resolutions considered here, the encoder-decoder architecture (purple crosses) achieves high performance with a bottleneck whose dimension is only two. This suggests that the high-resolution anomaly field can be represented by a bidimensional (latent) variable, which we denote $\boldsymbol{\mathrm{org}} $, for the purpose of modeling precipitation. $\boldsymbol{\mathrm{org}} $ encapsulates the aspects of convective organization that explain why precipitation can be different (stochastic) for the same low-resolution inputs \citep{moseley2016intensification}. This simplifies the question to how well we can represent $\boldsymbol{\mathrm{org}} $ from the time series of low-resolution inputs. The encouraging results of our models leveraging temporal memory, along with the finding of \cite{shamekh2023implicit} that $\boldsymbol{\mathrm{org}} $ is accurately predicted by an autoregressive model informed by the coarse-scale environment's past timesteps, suggest a mostly positive response. This reassuringly confirms that Earth system models with limited spatial resolutions can realistically represent coarse-scale precipitation as long as efforts to improve precipitation parameterization continue \citep{schneider2024opinion}. In hybrid modeling applications, where the NN itself computes precipitation at previous time steps, error propagation may reduce the informativeness of temporal memory---a limitation worth exploring in future work. More broadly, our results confirm that Pareto-optimal model hierarchies are useful in empirically establishing the partial equivalence between temporal memory and spatial granularity. This has practical applications for multi-scale dynamical systems that are not self-similar, where ergodicity cannot be used to deterministically infer detailed spatial information from coarse time series data.

\section{Conclusion \label{sec:Conclusion}}

In this study, we demonstrated that Pareto-optimal model hierarchies within a well defined (complexity, error) plane not only guide the development of data-driven models—from simple equations to sophisticated neural networks—but also promote process understanding. While these hierarchies are derived empirically based on the models and data at hand, they serve as practical approximations for exploring trade-offs between error and complexity. To distill knowledge from these hierarchies, we propose a multi-step approach: (1) use models along the Pareto front to hierarchically investigate the added value that leads to incremental error decreases from the simplest to the most complex Pareto-optimal model; (2) generate hypotheses and propose models tailored to the system of interest based on this added value; and (3) once the models are sparse enough to be interpretable, reveal the system's underlying properties that previous theories or models may have overlooked. Beyond knowledge discovery, such hierarchies promote interpretability by explaining the added value of models that may initially seem overly complex, and sustainability by optimizing models' computational costs once their added value is justified.

We have chosen three weather and climate applications in realistic geography settings to showcase the potential of machine learning in bridging fundamental scientific discovery and increasing operational requirements. Each case demonstrates a different nature of discovery: data-driven equation discovery for cloud cover parameterization, physics-guided architectures, informed by the available Pareto-optimal solution, that reflect the bidirectionality and vertical invariance of shortwave radiative transfer, and spatial information hidden within time series of the coarse environment when diagnosing tropical precipitation. However, all three insights required retaining the full family of trained models and might have been overlooked had we focused on a single optimal model.

In all three cases, neural networks proved particularly advantageous as they can quickly explore large datasets and generate hypotheses about problems that are or appear high-dimensional. We nonetheless emphasize that within this framework, deep learning is an integral part but not the ultimate goal of the knowledge generation process. Simpler models rivaling the accuracy of deep neural networks, initially intractable, may emerge once the necessary functional representations, features, and spatio-temporal connectivities are distilled. From this perspective, a rekindled interest in multi-objective optimization and hierarchical thinking would open the door to extracting new, human-interpretable scientific knowledge from ever-expanding geoscientific data, while paving the way for the adoption of machine learning in operational applications by fostering informed trust in their predictions.

\acknowledgments
T. Beucler acknowledges partial funding from the Swiss State Secretariat for Education, Research and Innovation (SERI) for the Horizon Europe project AI4PEX (Grant agreement ID: 101137682). A. Grundner acknowledges funding by the European Research Council (ERC)
Synergy Grant ``Understanding and Modeling the Earth System with Machine Learning (USMILE)'' under the Horizon 2020 research and innovation programme (Grant agreement No. 855187).  R. Lagerquist acknowledges support from the NOAA Global Systems Laboratory, Cooperative Institute for Research in the Atmosphere, and NOAA Award Number NA19OAR4320073. S. Shamekh acknowledges support provided by Schmidt Sciences, LLC, and NSF funding from the LEAP Science and technology center grant (2019625-STC).

%
%
\datastatement
The reduced data and code necessary to reproduce this manuscript's figures are available in the following GitHub repository: \url{https://github.com/tbeucler/2024_Pareto_Distillation}. The release for this manuscript is archived via Zenodo using the following DOI: \url{https://zenodo.org/records/13217736}. The cloud cover schemes and analysis code can be found at \url{https://github.com/EyringMLClimateGroup/grundner23james_EquationDiscovery_CloudCover/tree/main} and are preserved at \url{https://doi.org/10.5281/zenodo.8220333}. The coarse‐grained model output used to train and evaluate the data-driven models amounts to several TB and can be reconstructed with the scripts provided in the GitHub repository. This work used resources of the Deutsches Klimarechenzentrum (DKRZ) granted by
its Scientific Steering Committee (WLA) under project ID bd1179.  For radiative transfer, we use the ml4rt Python library (\url{https://github.com/thunderhoser/ml4rt}), with version 4.0.1 preserved at \url{https://doi.org/10.5281/zenodo.13160776}). The dataset for radiative transfer is stored at \url{https://doi.org/10.5281/zenodo.13159877}. The BRNN architecture used can be created with the file \texttt{peter\_brnn\_architecture.py}; all other architectures can be created with the scripts named \texttt{pareto2024\_make\_*\_templates.py}. The precipitation example uses "Precip-org" (\url{https://github.com/Sshamekh/Precip-org}), a Python repository managed by Sara Shamekh. DYAMOND data management was provided by the DKRZ and supported through the projects ESiWACE and ESiWACE2. The full data is available on the DKRZ HPC system through the DYAMOND initiative (\url{https://www.esiwace.eu/services/dyamond-initiative}).

%

\appendix[A]
\appendixtitle{Calibrated Parameters for Cloud Cover Parameterization}

In this appendix, we provide the values of the calibrated parameters in this manuscript's equations. 

For the Sundqvist scheme (Eq.~\ref{eq:C_Sundqvist}): $ \mathrm{RH}_{\text{surf}} = 0.55 $, $ \mathrm{RH}_{\text{top}} = 0.01 $, $ \mathrm{RH}_{\text{sat}} = 0.9/0.95 $ (over land/sea), and $ n = 2.12 $. 

For the two-feature linear regression (Eq.~\ref{eq:2F}): $ \alpha_{2F} = 1.027 $, $ \mathrm{RH}_{2F} = 0.827 $, and $ \Delta T_{2F} = 198.6 \text{K} $. 

For the four-feature linear regression (Eq.~\ref{eq:Multilinear}): $ \alpha_{1} = 1.861 $, $ \mathrm{RH}_{1} = 1.898 $, $ \Delta T_{1} = 265.9\text{K} $, $ q_{1} = 0.3775\text{g/kg} $, $ H_{1} = 104.3\text{m} $. 

For the quadratic model (Eq.~\ref{eq:Quadratic}): $ \alpha_{2} = 2.070 $, $ \mathrm{RH}_{2,1} = 2.021 $, $ \Delta T_{2,1} = 259.7\text{K} $, $ q_{2,1} = 0.4741\text{g/kg} $, $ q_{2,2} = 3.161\text{g/kg} $, $ q_{2,3} = 0.1034\text{g/kg} $, $ H_{2} = 665.6\text{m} $, $ \mathrm{RH}_{2,2} = 2.428 $, $ \Delta T_{2,2} = 206.0\text{K} $.

For the Wang scheme: $ \alpha_{\mathrm{Wang}} = 0.9105 $, $ \beta_{\mathrm{Wang}} = 914 \cdot 10^{3} $.

For the PySR equation, $\overline{{\cal C}}=0.4435$,  
\begin{equation}
    \left(\frac{\partial{\cal C}}{\partial\mathrm{RH}}\right)_{\overline{\mathrm{RH}},\overline{T}} = 1.159, \left(\frac{\partial{\cal C}}{\partial T}\right)_{\overline{\mathrm{RH}},\overline{T}} = \frac{0.0145}{\text{K}} ,
\end{equation}
\begin{equation}
    \left(\frac{\partial^{2}{\cal C}}{\partial\mathrm{RH}^{2}}\right)_{\overline{\mathrm{RH}},\overline{T}} = 4.06, \left(\frac{\partial{\cal C}}{\partial\mathrm{RH}\partial T^{2}}\right)_{\overline{\mathrm{RH}},\overline{T}} = \frac{1.32.10^{-3}}{\text{K}^2},
\end{equation}
$ H_{\mathrm{PySR}} = 585 \text{m} $, $ \left(d\mathrm{RH}/{dz}\right)_{\mathrm{max\ {\cal C}}} = -2/\text{km} $, $ \epsilon_{\mathrm{PySR}} = 1.06 $, $ \lambda_{\ell} = 3.845.10^5 $, $ \lambda_{i} = 1.448.10^6 $.

For the GP-GOMEA equation: $ \alpha_{G} = 66 $, $ \beta_{G} = 1.36.10^{-4} $, $ \mathrm{RH_{G}} = 11.5\% $, $ \gamma_{G} = 0.194 $, $ q_{G,i} = 4.367\text{mg/kg} $, $ \delta_{G} = 0.774 $, $ q_{G,\ell+} = 88.05\text{mg/kg} $, $ q_{G,\ell-} = 5.61\text{mg/kg} $, and $ \epsilon_{\mathrm{G}} \rightarrow 0$ (0 cloud cover assigned in the absence of condensates and model calibrated only with condensates present).  


\appendix[B]
\appendixtitle{Features for Shortwave Radiative Transfer Emulation}

The 22 features $\boldsymbol{X_\eta}$ with a vertical profile are temperature $\boldsymbol{T}$ [K], pressure $\boldsymbol{p}$ [Pa], specific humidity $\boldsymbol{q}$ [kg kg$^{-1}$], relative humidity $\boldsymbol{\mathrm{RH}}$ [\%], liquid water content $\boldsymbol{\mathrm{LWC}}$ [kg m$^{-3}$], ice water content $\boldsymbol{\mathrm{IWC}}$ [kg m$^{-3}$], downward and upward liquid water path $\boldsymbol{\mathrm{LWP}}_{\downarrow},\boldsymbol{\mathrm{LWP}}_{\uparrow}$ [kg m$^{-2}$], downward and upward ice water path $\boldsymbol{\mathrm{IWP}}_{\downarrow},\boldsymbol{\mathrm{IWP}}_{\uparrow}$ [kg m$^{-2}$], downward and upward water vapor path $\boldsymbol{\mathrm{WVP}_{\downarrow}},\boldsymbol{\mathrm{WVP}_{\uparrow}}$ [kg m$^{-2}$], O$_3$ mixing ratio [kg kg$^{-1}$], height above ground level $\boldsymbol{z}$ [m], height layer thickness $\boldsymbol{\Delta z}$ [m], pressure layer thickness $\boldsymbol{\Delta p}$ [Pa], liquid effective radius $\boldsymbol{r_\ell}$ [m], ice effective radius $\boldsymbol{r_i}$ [m], N$_2$O concentration [ppmv], CH$_4$ concentration [ppmv], CO$_2$ concentration [ppmv], and aerosol extinction [m\textsuperscript{-1}].





%



\bibliographystyle{ametsocV6}
\bibliography{Refs}

\end{document}